\documentclass[twocolumn,showpacs,preprintnumbers,amsmath,amssymb,pra]{revtex4}
\usepackage{graphicx}
\usepackage{dcolumn}
\usepackage{bm}
\begin{document}
 \preprint{version 5}

\title{2S hyperfine structure of atomic deuterium}

\author{N. Kolachevsky}
\altaffiliation[Also at P.N. Lebedev Physics Institute, Moscow,
Russia] \
\author{P. Fendel}
\author{S.G. Karshenboim}
\altaffiliation[Also at D.I. Mendeleev Institute for Metrology,
St. Petersburg, Russia] \
\author{T.W. H\"{a}nsch}
\altaffiliation[Also at Ludwig-Maximilians-University, Munich,
Germany] \ \affiliation{Max-Planck-Institut f\"{u}r Quantenoptik,
85748 Garching, Germany}

\date{\today}

\begin{abstract}
We have measured the frequency splitting between the $(2S, F=1/2)$
and $(2S, F=3/2)$ hyperfine sublevels in atomic deuterium by an
optical differential method based on two-photon Doppler-free
spectroscopy on a cold atomic beam. The result $f_{\rm
HFS}^{(D)}(2S)= 40\, 924\, 454(7)$~Hz is the  most precise value
for this interval to date. In comparison  to the previous
radio-frequency measurement we have improved the accuracy by the
factor of three.
 The specific combination
of hyperfine frequency intervals for metastable- and ground states
in deuterium atom $D_{21}=8f_{\rm HFS}^{(D)}(2S)-f_{\rm
HFS}^{(D)}(1S)$ derived from our measurement is in a good
agreement with $D_{21}$ calculated from quantum-electrodynamics
theory.
\pacs {12.20.Fv, 32.10.Fn, 32.30.Jc, 42.62.Fi}
\end{abstract}

\maketitle
\section{introduction}

High-precision measurements in light atomic systems permit
accurate tests of the  quantum-electrodynamics theory (QED). QED
calculations enter into a number of fundamental values related to
free particles and simple atoms. In conventional atomic systems
the accuracy of QED tests is restricted by insufficient knowledge
of the nuclear structure which is the main obstacle on the way to
improve theoretical predictions for the Lamb shift and the
hyperfine structure in hydrogen (see e.g. \cite{Eides,sgkpsas}).

The leading corrections for different energy levels coming from
nuclear size effects are proportional to the squared value of
non-relativistic wave function of the electron at the origin:
\begin{equation}
\Delta E_{\rm{nucl}}=A_N\cdot|\psi(r=0)|^2\;,
\end{equation}
where the value of the coefficient $A_N$ is determined by
parameters of the nucleus and does not depend on atomic quantum
numbers. For $s$--levels in hydrogen-like systems the
non-relativistic wave function at the origin scales with the
principal quantum number $n$ as $|\psi(r=0)|^2\sim1/n^3$. Thus, if
one takes the following difference of the energies $E(nS)$
corresponding to two different $s$--levels $n'^3 E(n'S)-n^3
E(nS)$, the leading contribution of nuclear effects cancels.

Recently, a significant progress in calculations of the $8 E_{\rm
HFS}(2S)-E_{\rm HFS}(1S)$ difference of $2S$ and $1S$ hyperfine
splitting (HFS) intervals in light hydrogen-like atoms has been
achieved \cite{Karsh1}. New state-dependent QED terms to the HFS
interval frequency $f_{\rm HFS}(nS)$ up to the fractional order of
$\alpha^4$ and $\alpha^3\,m_e/m_p$ have been calculated as well as
the next-to-leading nuclear structure effects. The
 accuracy of this theoretical prediction for $D_{21}$ exceeds now
 the experimental accuracy which is mainly restricted by the uncertainty in
 the
 determination of the $f_{\rm HFS}(2S)$.

Experimental results are available for the $1S$ and $2S$ hyperfine
interval in hydrogen \cite{Ramsey,Kusch, Hessels,Kolachevsky},
deuterium \cite{Wineland, Reich} and the helium-3 ion
\cite{Schl69,Prior}.  The study of neutral atoms and ions requires
different experimental techniques. While $1S$ HFS intervals in
neutral atoms \cite{Ramsey,Wineland} have been measured with a
higher relative accuracy as in the helium-3 ion \cite{Schl69}, the
$2S$ HFS intervals in hydrogen  \cite{Hessels, Kolachevsky} and
deuterium \cite{Reich} are known less precisely as in the helium-3
ion \cite{Prior}. Since the traditional microwave methods have
likely reached their limits we have been working on an optical
determination of the hyperfine interval in the metastable $2S$
state of hydrogen and deuterium.

Recently we have measured the hyperfine splitting of the $2S$
state of the hydrogen atom applying an optical technique
\cite{Kolachevsky}. It is based on two-photon spectroscopy on a
cold beam shielded from magnetic fields. The $2S$ HFS interval has
been determined from the frequency difference of two stable light
fields exciting $1S$--$2S$ transitions for singlet ($F=0$) and
triplet ($F=1$) components. The differential method cancels a
number of systematic effects intrinsic to two-photon spectroscopy
which provides a significant increase of accuracy in comparison
with absolute frequency measurements \cite{Niering, Hbook,
Fischer}.

We have improved this technique \cite{Kolachevsky} and applied it
to the spectroscopy of atomic deuterium. The hyperfine structure
in deuterium is approximately 4 times smaller than in hydrogen and
its accurate optical measurement is a harder problem, however, a
smaller HFS value has also some advantages due to a reduction of
certain systematic effects for the comparison of the $1S$--$2S$
frequencies between different spin states. The experiment and its
systematic effects are presented in two following sections while a
comparison of theory and experiment is summarized in the
concluding part of the paper.

\section{Measurement of the $2S$ hyperfine splitting}

The $2S$ HFS interval has been measured once  by Reich, Heberle,
and Kusch \cite{Reich}. The value of
\begin{equation}
 f_{\rm
HFS}^{(D)}(2S)= 40\, 924\, 439(20)~\rm{Hz}
\end{equation}
has been obtained by a radio-frequency (rf) method applied earlier
to a similar measurement in hydrogen \cite{Kusch}. In comparison
to the hydrogen measurement \cite{Kusch}, the absolute accuracy of
determination of $f_{\rm HFS}^{(D)}(2S)$ \cite{Reich} has been
improved by a factor of 3. In spite of the fact that the HFS
intervals in deuterium are strongly affected by magnetic fields,
as in hydrogen, the average of two rf transitions frequencies in
the metastable deuterium atom $(F=1/2,
m_F=-1/2)\leftrightarrow(F=3/2, m_F=1/2)$ and $(F=1/2,
m_F=1/2)\leftrightarrow(F=3/2, m_F=-1/2)$ contains no linear
field-dependent terms and thus is rather insensitive to the field.
Still, the most important systematic effects contributing to the
20 Hz error were the uncertainty in the determination of the
magnetic field and the rf Stark effect.

\begin{figure}[t]
\begin{center}
\includegraphics [width=50 mm]{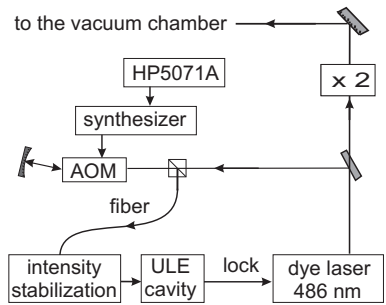}
\caption{Laser system for $1S$--$2S$ deuterium
spectroscopy.}\label{fig1a}
\end{center}
\end{figure}

For an independent measurement of $f_{\rm HFS}^{(D)}(2S)$ we use
the hydrogen spectrometer setup described elsewhere
\cite{Fischercan}. A dye laser operating near 486 nm is locked to
a definite TEM$_{00}$ mode of an ultra-stable cavity with
Ultra-Low Expansion (ULE) glass spacer and a drift less than 0.5
Hz/s (Fig.\ref{fig1a}). To change the laser frequency with respect
to the cavity mode we use a broadband double-pass acousto-optic
modulator (AOM) placed between the laser and the cavity. The
frequency of the dye laser is doubled in a Barium $\beta$-borate
crystal, and the resulting 243 nm radiation is coupled to a linear
enhancement cavity inside the vacuum chamber where  the two-photon
excitation takes place.

\begin{figure}[t]
\begin{center}
\includegraphics [width=85 mm]{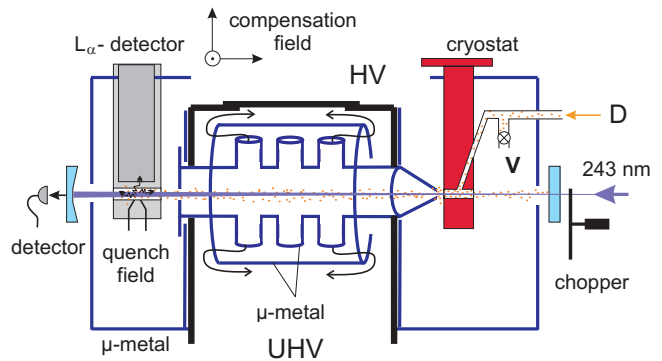}
\caption{Vacuum part of the experiment. HV and UHV are the high-
and ultrahigh vacuum regions, while V is the bypass valve reducing
the atomic flow escaping from the nozzle.}\label{fig2}
\end{center}
\end{figure}

Atomic deuterium produced in a 15 W, 2.5 GHz rf gas discharge
flows through teflon tubes to a copper nozzle cooled by a
flow-through cryostat (Fig.\ref{fig2}). Atoms thermalize with the
nozzle walls and leave the nozzle in both directions coaxially
with the enhancement cavity mode. A part of the atoms flying along
the mode are excited to the $2S$ state and then reach the
detection zone where a weak electrical quenching field is applied.
The field mixes the $2S$ and $2P$ states of the atoms causing
their fast decay to the ground state with the emission of
Lyman--$\alpha$ photons. Photons are counted by a solar-blind
photomultiplier connected to a photon counter incorporated into
time-of-flight measurement scheme. The excitation radiation is
periodically blocked before the cavity by a chopper wheel, and 121
nm photons are detected only within a 3 ms dark time interval.
Introducing a delay $\tau$ between closing the beam and starting
the detection, we can thus select different velocity groups
contributing to the signal. The delay $\tau$ sets an upper limit
for the atomic velocity to $v_{\rm{max}}= l/\tau$, where $l$ is
the distance between the nozzle and detector. Using a
multi-channel scaler we sort the counts in adjacent time bins
($\tau=10\,\mu s,\ 210\,\mu s,\ 410\,\mu s,\ldots$,) and
simultaneously record  up to 12 spectra containing the information
about velocity-dependent effects, e.g. 2nd order Doppler effect.

Here we will point out some important changes which have been
introduced in the spectrometer configuration (which has been used
for the recent measurements in atomic hydrogen \cite{Kolachevsky,
Fischer}) to optimize it for the current deuterium measurement. To
maximize the count rate for slow atoms ($v\sim200$ m/s) one has to
find a compromise between thermalization and recombination
processes at the cold walls of the nozzle by adjusting the nozzle
geometry and temperature. The best rates are observed with a
larger nozzle diameter (2--2.5 mm compared to 1.2 mm for hydrogen)
and at slightly higher temperatures (6--6.5 K compared to 5 K).
Typically, the count rate with the deuterium beam has been higher
than for hydrogen  (by a factor 2--5 depending on $\tau$).

On the way between the gas discharge and the nozzle we have
installed a bypass valve which can be opened to a high-vacuum (HV)
part of the vacuum chamber, where the cavity mirrors, the nozzle
and the detector have been placed. By opening the valve we reduce
the atomic flow through the nozzle by a factor of 4. Even at the
lowest pressure in the gas discharge (0.7 mbar) and with opened
valve, the count rate was sufficient to detect a solid $2S$ signal
for the atoms with $v\sim200$ m/s. The high-vacuum volume is
pumped by a turbo-molecular pump to  $5\cdot10^{-5}$ mbar.

Atoms are collimated by two 1.5 mm diaphragms separating the HV
and the ultrahigh-vacuum (UHV) volumes. Most of the interaction
region (95\%) is in the UHV region pumped by a $10^4$ l/s
cryopump. Without the atomic beam, the background gas pressure in
the  UHV region equals $3\cdot10^{-8}$ mbar. The background gas
pressure increases with the deuterium flow escaping from the
nozzle and can be thus varied up to $2\cdot10^{-7}$ mbar.

The excitation region is shielded against external magnetic
fields. As in \cite{Kolachevsky}, we use a two-stage $\mu$--metal
shielding. The outer shielding together with the compensation
field reduces residual fields over the entire interaction region
 to 10--20 mG. Configuration of the inner shielding has been
improved to increase the shielding factor as well as its
throughput for pumping.

\begin{figure}[t]
\begin{center}
\includegraphics [width=50 mm]{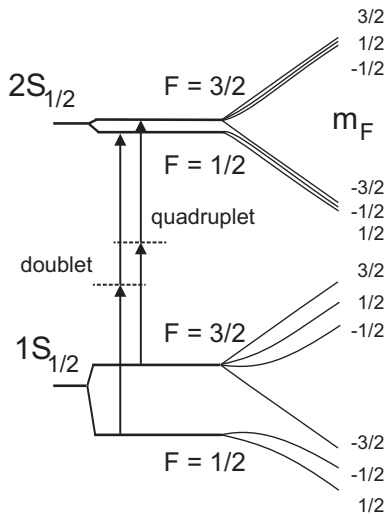}
\caption{Two-photon transitions between hyperfine components of
the $1S$ and $2S$ levels in atomic deuterium. The splitting of
magnetic sublevels in a magnetic field is also
presented.}\label{fig1}
\end{center}
\end{figure}

In the absence of magnetic fields, the $2S$ HFS interval frequency
is given by the following combination of optical frequencies and
the ground state HFS interval:
\begin{equation}
f_{\rm HFS}^{(D)}({2S}) = f_{\rm
HFS}^{(D)}({1S})+f_{F=3/2}-f_{F=1/2}, \label{neq1}
\end{equation}
where $f_{F=1/2}$ and $f_{F=3/2}$ are the frequencies of the
doublet and the quadruplet transitions at 121 nm (see
Fig.\ref{fig1}).  In our experiment, we measure the frequency
difference $(f_{F=3/2}-f_{F=1/2})$ using  two-photon spectroscopy
of the corresponding transitions. The frequency of the
ground-state splitting has been measured by  Wineland and Ramsey
\cite{Wineland} with an uncertainty of $5\cdot10^{-12}$:
\begin{equation}
f_{\rm{HFS}}^{(D)}(1S)=327\,384\,352.5222(17)\ \rm{Hz}.
\end{equation}
The contribution to the resulting error budget  introduced by this
uncertainty is negligible.

For allowed two-photon transitions between the levels with $\Delta
F=0$, the contribution of the linear Zeeman effect cancels and the
HFS frequency shift in a magnetic field $B$  scales as $B^2/f_{\rm
HFS}$ which is unfavorable for deuterium in comparison to
hydrogen. But due to a low-field regime and the absence of linear
terms, our measurement is rather insensitive to magnetic fields.
In our case, the magnetic field $B$ shifts the frequency $f_{\rm
HFS}^{(D)}({2S})$ approximately as $35\cdot B^2$ kHz/G$^2$.

\begin{figure}[t]
\begin{center}
\includegraphics [width=60 mm]{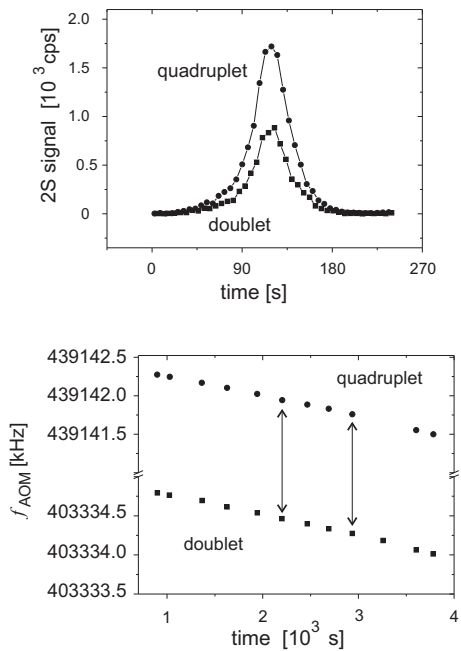}
\caption{Simultaneous recording of the two transition lines (top).
AOM frequencies $f_{\rm{AOM}}$ corresponding to the centers of the
doublet and quadruplet transition lines (bootom). The frequency
change is due to the drift of the reference ULE cavity (bottom).
}\label{fig3}
\end{center}
\end{figure}

The laser as well as the lock electronic are adjusted in such a
way, that the two $(1S, F=1/2)\rightarrow (2S, F=1/2)$ (doublet)
$(1S, F=3/2)\rightarrow (2S, F=3/2)$ (quadruplet) transitions can
be excited by changing the AOM frequency (see Fig.\ref{fig1}).

 To measure the
frequency difference between the two  two-photon transitions we
have applied the following procedure.  The frequency of the
double-pass AOM placed between the reference ULE cavity and the
laser is tuned to one of the transitions  and one point of the
spectrum is measured with a photon accumulation time of 0.5 s.
After the measurement,  the AOM frequency $f_{\rm AOM}$ is sweeped
over a big
 fixed frequency interval (approximately equals to $[f_{\rm
HFS}^{(D)}({1S})-f_{\rm HFS}^{(D)}({2S})]/8\approx36$ MHz) to
reach the other two-photon transition.  During the sweep, the
laser has been continuously kept in lock. The factor 8 arises from
the double-pass scheme for the AOM, the second harmonic generation
stage and  the two-photon process. Than we introduce 0.5 s pause
as recovery time for  multiple feedbacks, and repeat the
measurement of one point of the other two-photon transition. After
the measurement we add a small frequency step for scan and sweep
the synthesizer back to the first transition. The result of such a
procedure is presented on Fig.\ref{fig3}(top).

The rf synthesizer driving the AOM is continuously referenced to
the 10 MHz signal of a HP5071A cesium frequency standard
calibrated by GPS (Fig.\ref{fig1a}). The
 miscalibration of the primary standard introduces a negligible uncertainty to
the measured difference. After the double-pass AOM, the  laser
light  is spatially filtered by means of a single-mode fiber,
while the intensity of the light coupled into the cavity is
actively stabilized. That guarantees that the coupling conditions
which can change the frequency of the reference cavity mode
\cite{Nevsky} are the same for both transitions.

Both simultaneously recorded doublet and quadruplet transition
lines are fitted with a Lorentzian function in the time and
frequency domains. A typical time interval between the centers of
such a line pair
 is less than 5 s. This corresponds to some small
correction on a level of a few hertz which should be introduced to
correct for the reference cavity drift (fig.\ref{fig3}(bottom)).
Such a procedure significantly reduces the influence of reference
cavity frequency fluctuations on the $2S$ HFS data statistics. The
new method suppresses the influence coming from all drifts in the
experimental setup with a time scale exceeding 200 s which is
required to record one line. Even though such fluctuations do not
introduce systematic shifts, a long time  is required to average
them. The approach is much more forgiving to systematic effects:
one independent point is obtained after 3 minutes of measurement,
while for the previous method \cite{Kolachevsky} the corresponding
time interval was  about  20 minutes.

\section{Results and systematic effects}

We have measured $f_{\rm HFS}^{(D)}({2S})$ during 7 days using the
new method and during 6 days applying the method described in
\cite{Kolachevsky}. To test for systematic effects we have varied
different parameters of the experiment within each day. We have
recorded over 1000 deuterium time-resolved spectra. The averaged
amplitude ratio between quadruplet and doublet transitions equals
to 2.00(2) which is in a good agreement with  theoretical
expectation.

The systematic effects for our two-photon beam spectroscopy are
well-known \cite{Hbook}. For the differential measurement, the
most
 significant  effects cancel out. As
shown in \cite{Kolachevsky}, the differential dynamic Stark shift
cancels to the level of $10^{-6}$ relative to the shift of the
$1S$--$2S$ transition (about 500 Hz). Because of some residual
fluctuations of the 243 nm radiation intensity, the doublet and
quadruplet lines acquire slightly different shifts. We correct for
this difference for each line by monitoring the power leaking out
of the cavity (Fig.\ref{fig2}). The resulting correction for the
new method equals 0.5 Hz which is a factor 4 lower than for the
measurement in hydrogen \cite{Kolachevsky}. For the simultaneous
recording technique, long-term intensity fluctuations in the
enhancement cavity contribute less to the result than in the case
of \cite{Kolachevsky}. Besides the correction, we add a 0.5 Hz
uncertainty to the error budget.

A DC electrical field $E$ mixes $2S$ and $2P$ levels causing a
$2S$ level shift. According to \cite{Bethe}, the energy shift of
the $2S$ level is reversely proportional to the Lamb shift. Thus,
the differential shift  scales as $E_{\rm{HFS}}/L_{2S-2P}$, where
$L_{2S-2P}$ is the Lamb shift of the $2S$ level. The evaluated
shift of the $2S$ HFS interval in deuterium equals $-300$ $E^2$ Hz
cm$^2$/V$^2$ and is about 4 times smaller as in hydrogen. The
excitation region is shielded from stray fields by coating all
surrounding parts of the interaction region with graphite.
Residual stray fields in the setup are estimated to be less than
30 mV/cm \cite{Huber}, which corresponds to a $2S$ HFS frequency
shift on a level of $-0.3$ Hz. We  add an uncertainty of 0.5 Hz to
the error budget considering the slightly worse geometrical
properties of our shielding compared to the Faraday cage used in
\cite{Huber}.

Residual magnetic fields split the magnetic sublevels and shift
the measured hyperfine splitting to higher values. The shift
mostly originates from imperfectly shielded parts of the
excitation region lying in the HV region (Fig.\ref{fig2}). To
check for this effect we have once made a set of measurements
without compensation field and observed a change of $f_{\rm
{HFS}}^{(D)}(2S)$ equal to 20(22) Hz. Without compensation field,
the measured value of
 magnetic fields around the nozzle and the detector is
about 200 mG.  The compensation field reduces the  magnetic fields
for at least a factor of 10, which means that the residual shift
should be on a sub-hertz level. We conservatively estimate the
shift as 0.5(1.0) Hz.

Another source for level shifts is a pressure shift caused by
collisions with the background gas and within the beam itself. The
ground-state hyperfine splitting in hydrogen and deuterium  is
rather insensitive to collisions. Typically, this shift is on a
level of 1 Hz/mbar depending on the buffer gas \cite{Anderson,
Morgan} which corresponds to a vanishing shift for our pressure
range. On the other side, there are no reliable experimental data
for the pressure shift of the $2S$ HFS frequency \cite{Hessels,
Kolachevsky}. The upper limit for the $2S$ HFS frequency shift can
be taken as the frequency shift of the
$1S\,(F=1,m_F=\pm1)\leftrightarrow2S\,(F'=1,m'_F=\pm1)$ transition
 equal to $-8(2)$ MHz/mbar \cite{McIntyre, Kleppner}. But considering the theoretical works \cite{Davison, Ray},
 there is no reason to expect the $2S$ HFS shift to be orders of
 magnitude higher than for the ground state.

\begin{figure}[t]
\begin{center}
\includegraphics [width=75 mm]{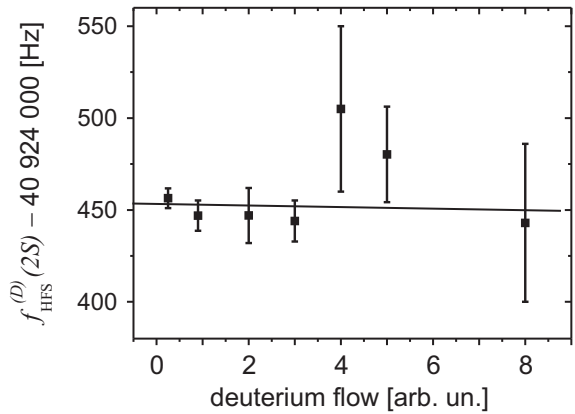}
\caption{The $2S$ HFS frequency  vs. deuterium flow. One unit
corresponds to $3.6\cdot10^{17}$ particles coming to the nozzle
per second. The data is obtained using the simultaneous recording
method and corresponds to the delay time of $ \tau=410\ \mu$s. To
extrapolate to the zero flow we fit the data with a linear
function (solid line).}\label{fig4}
\end{center}
\end{figure}

\begin{table} [b]
\caption{$2S$ HFS frequency and statistical uncertainty of
extrapolation for different delays $\tau$. The velocity
$v_{\rm{max}}$ represents the maximal velocity of atoms
contributing to the signal. The Coefficient $k$ is the slope of
the linear fit (Fig.\ref{fig4}).}
\begin{tabular}{c @{\ \ \ \ }c  @{\ \ \ \ } c @{\ \ \ \ } c}

  \hline
  \hline
  \rule{0pt}{3ex}
 $\Delta\tau$, $[\mu$s]   & $v_{\rm{max}}$, [m/s] & $f_{\rm {HFS}}^{(D)}(2S)$, [Hz]& $k$, [Hz/a.u.] \\
  \hline
  \rule{-5pt}{3ex}
10&$-$&40\,924\,452(7)&4(4)\\
210&1000&40\,924\,454(3)&0.6(2.0)\\
410&510&40\,924\,453(5)&$-$0.4(3.0)\\
610&340&40\,924\,448(10)&0.3(6.0)\\
810&260&40\,924\,446(15)&$-$5(9)\\
 \hline \hline
\end{tabular}
\label{tabl1}
\end{table}

 We have experimentally investigated the influence of collisions
 on the $f_{\rm {HFS}}^{(D)}(2S)$ frequency. Most of the collisions take
 place within the atomic beam where the pressure can be much higher than the
 background gas pressure. Using the  method of simultaneous detection, we
 have varied the atomic flow over a wide range by changing the
 pressure in the gas discharge (between 0.8 mbar and 8 mbar) and
 by opening the bypass valve (Fig.\ref{fig2}). The background gas pressure
  scales linearly with the flow with a small offset of
  $3\cdot10^{-8}$ mbar. Experimental data are presented in
  Fig.\ref{fig4}. Each point represents the result of statistical averaging of
  multiple data points detected at different days. About 1/4 of all the data have been taken at the lowest pressure in
  the gas discharge with an
  opened valve (the left point). At high flows the slow atoms are accelerated and pushed away from the beam by collisions
(Zacharias effect)  \cite{Huber} which causes a loss in statistics
and corresponding increase of uncertainty. Moreover, the
measurement time
  becomes restricted by  a fast growth of a film of molecular deuterium
  on the nozzle. For the previous method \cite{Kolachevsky} the
  flow   range of $\{2$\,-\,$8\}$ (Fig.\ref{fig4}) is practically unreachable.
  Disregarding the delay time ${\tau}$ we have not observed
  any indication for a $2S$ HFS interval pressure shift on our level of accuracy.

\begin{figure}[t]
\begin{center}
\includegraphics [width=80 mm]{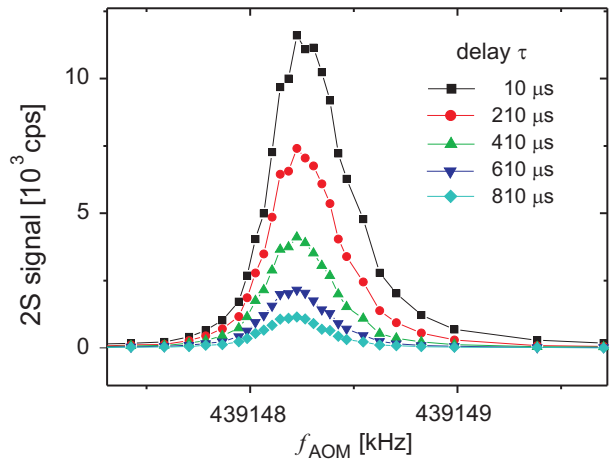}
\caption{Quadruplet transition lines for the different delays
$\tau$.}\label{fig6}
\end{center}
\end{figure}

The observed transition lines  are shifted by the second order
Doppler effect in a range between 0.1--1 kHz depending on the
velocity distribution of the atoms contributing to the signal
\cite{Hbook}. The lines are not symmetric and significantly differ
from Lorentzian profiles at short ($\tau<410 \mu s$) delays
(Fig.\ref{fig6}). If we excite two-photon transitions in the same
thermal beam and detect the signal with some precisely defined
delay time $\tau$ equal for both singlet and quadruplet
transitions, the shift and the influence of the asymmetry should
cancel out. To estimate the possible uncertainty coming form the
"wrong" fitting procedure with a Lorentzian function and all other
velocity-dependent effects, we have evaluated the $f_{\rm
{HFS}}^{(D)}(2S)$ frequency for different delays. The data for
each delay has been corrected for the AC Stark shift, and then the
extrapolation to  zero flow has been made (Fig.\ref{fig4}). The
results of this extrapolation  are presented in the Table
\ref{tabl1}. All velocity-dependent systematic effects, if they
exist, should reveal themselves in such evaluation procedure.

All the results are consistent within the error bars. For higher
delays $\tau$ the corresponding uncertainty grows due to the lack
of statistics. As a final value we have chosen the result $f_{\rm
{HFS}}^{(D)}(2S)= 40\,924\,453(5) \  {\rm Hz} $
 ($\tau=410$ $\mu$s) due to the following reasons: (i) the statistical
 uncertainty for this result is small, (ii) the line-profile is
 practically undistinguishable from Lorentzian, and (iii) the fast
 non-thermalized
 atoms do not contribute to the signal. We add an additional
uncertainty of 0.3 Hz corresponding to the conservatively
estimated shift coming from the collisions with the residual
background gas.

The data obtained during 6 days of measurement applying the older
method presented in \cite{Kolachevsky} give the result of $f_{\rm
{HFS}}^{(D)}(2S)= 40\,924\,462(15)$ Hz. The data are measured only
at low deuterium flows $\{0.2$--$1.5\}$ and are simply averaged
over the entire ensemble. We have increased the uncertainty due to
the possible pressure shift (1 Hz) and the AC Stark shift (2 Hz).

Combining both these statistically independent results and adding
all known systematic uncertainties linearly, we arrive at the
final value for the $2S$ hyperfine interval in atomic deuterium
\begin{equation}
f_{\rm {HFS}}^{(D)}(2S)= 40\,924\,454(7) \  \rm{Hz}.
\end{equation}
This result is in a good agreement with the one obtained by an rf
method \cite{Reich}, but is more accurate by a factor 3. The
results as well as the error budget are collected in Table
\ref{tabl2}.

\begin{table} [t]
\caption{Summary of systematic errors and the final result for the
$2S$ hyperfine interval. The independent results obtained by the
"new" simultaneous recording method and the one from
\cite{Kolachevsky} are presented.}
\begin{tabular}{l @{\ \ \ \ }c  @{\ \ \ \ } c }

  \hline
  \hline
  \rule{0pt}{3ex}
 & Contribution [Hz] & Uncertainty [Hz] \\
 \hline
  \rule{-3pt}{3ex}
"New" method&40\,924\,453&5\\
 AC Stark shift&&0.5\\
 DC Stark shift&$-0.3$&0.5\\
Magnetic field&$0.5$&1.0\\
Pressure shift&&0.3\\
  \hline
    \rule{-3pt}{3ex}
Method \cite{Kolachevsky}&40\,924\,462&15\\
 AC Stark shift&&2\\
 DC Stark shift&$-0.3$&0.5\\
Magnetic field&$0.5$&1.0\\
Pressure shift&&1\\
\hline \rule{-5pt}{3ex}
Result&40\,924\,454&7\\
 \hline \hline
\end{tabular}
\label{tabl2}
\end{table}

\section{$D_{21}$ difference}

\begin{figure}[h!]
\begin{center}
\includegraphics [width=80 mm]{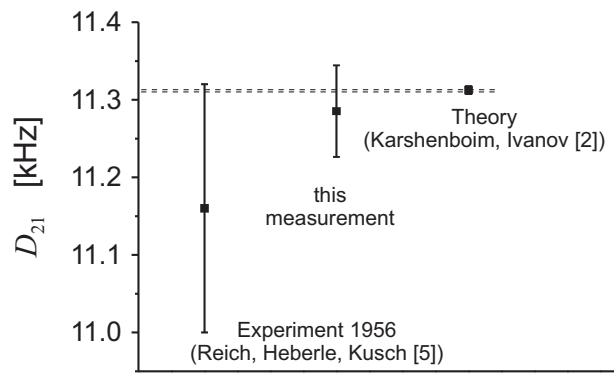}
\caption{Experimental and theoretical values for the $D_{21}$
values in deuterium. The dashed lines represent the uncertainty of
the theoretical value.}\label{fig5}
\end{center}
\end{figure}

To derive the $D_{21}$ difference we should combine our result
with the known value of the $1S$ HFS interval \cite{Wineland}:
\begin{equation}
D^{\rm
exp}_{21}=8f_{\rm{HFS}}^{(D)}(2S)-f_{\rm{HFS}}^{(D)}(1S)=11\,280(56)
\ \rm{Hz}.
\end{equation}
This result is in a good agreement with the theoretical prediction
\cite{Karsh1}
\begin{equation}
D^{\rm theor}_{21}=11\,312.5(5) \ \rm{Hz}.
\end{equation}
The experimental and theoretical values are presented in
Fig.\ref{fig5}. Restricted by the experimental accuracy, we can
test the state-dependent QED contributions on the level of 0.2 ppm
which can be compared to similar tests in hydrogen
\cite{Kolachevsky} and the $^3$He$^+$ ion \cite{Prior}. We point
out that the absolute accuracy of our measurement exceeds the both
cited results, while the relative accuracy suffers from the
relatively small HFS interval in deuterium.

Today precision optical methods challenge even in fields were
classical rf methods have been traditionally considered as
favorable. After a successful measurement of the Lamb shift
\cite{Hbook, Weitz} and the $2S$ the HFS interval
\cite{Kolachevsky} in the hydrogen atom we have now presented a
result on the $2S$ HFS interval in deuterium with an accuracy
exceeding that of rf methods.

The accuracy of the optical measurement can be improved by the
further reducing the atomic velocity. The current results indicate
that the pressure shift, which limits the accuracy of absolute
frequency measurements in ultra-cold hydrogen \cite{Kleppner}, may
not play such a crucial role for the differential technique.

The work was partially supported by DFG (grant \#
436RUS113/769/0-1) and RFBR (grant \#03-02-04029). The authors
wish to thank U. Jentschura, Th. Udem, and M. Fischer for useful
discussions.


\end{document}